\documentstyle[aps,preprint]{revtex}
\begin{document}

\draft
\preprint{IUCM95-033}
\title{Numerical Tests of the Chiral Luttinger Liquid Theory
for Fractional Hall Edges}
\author{J. J. Palacios and A. H. MacDonald}
\address{Department of Physics, Indiana University, Bloomington,
IN 47405, USA.}

\date{\today}
\maketitle

\begin{abstract}
We report on microscopic numerical studies which support the chiral Luttinger
liquid theory of the fractional Hall edge
proposed by Wen.  Our calculations are based in part on
newly proposed and accurate many-body trial wavefunctions for the low-energy
edge excitations of fractional incompressible states.

\end{abstract}


The quantum Hall effect (QHE) can occur only when a disorder-free
two-dimensional electron
system has an incompressibility, i.e., a discontinuity in the
chemical potential, at a magnetic field dependent density, $n^{*}(B)$.
The incompressibility implies a gap for charged and neutral
excitations in the bulk of the system, but the magnetic
field dependence of the density at which the gap occurs
requires the existence of gapless excitations localized at
the edge of the system.\cite{leshouches}
In equilibrium the current responsible for the orbital
diamagnetism of the underlying two-dimensional electron system
is carried at the edge and satisfies the thermodynamic identity,
\begin{equation}
\frac{\partial I}{\partial \mu} = c \frac{d n^{*}(B)}{d B}.
\label{eq:ahm1}
\end{equation}
In the {\it edge-state} picture,\cite{edgepicture}
the quantum Hall effect follows from Eq.(~\ref{eq:ahm1}) when
local equilibria are established on uncoupled edges.
For the case of bulk incompressibilities at fractional
Landau level filling factors ($\nu$), which are of many-body origin,
Wen's chiral Luttinger liquid (CLL) theory\cite{wen,wenreview}
of the low-energy edge excitations predicts non-Fermi-liquid
effects which have recently been of great interest.
\cite{wen,wenreview,kanefisher}
The simplest version of this theory, and the one we test
numerically here, applies to the edge of the incompressible ground states
which occur at  $\nu = 1/m$ for $m$ odd.\cite{Laughlin}

The non-Fermi-liquid properties of one-dimensional fermion systems
captured by the Luttinger model,\cite{mahanll} arise from
interactions between left-going and right-going particles in states
close to the  Fermi level.  In a single-particle model, left-going
and right-going states close to the Fermi
energy in the QHE regime are localized
on opposite edges of a sample with Hall bar geometry, and therefore interact
weakly.  However, Wen's theory predicts that in the fractional QHE regime,
non-Fermi-liquid behavior arises {\it without inter-edge interactions.}
We therefore consider a quantum Hall droplet (QHD) system for which
electrons are confined to
a finite area by a circularly symmetric external potential and for which
a single circular edge exists.
For such a system it follows\cite{wen,macdedge,macdjohnson} directly
from the microscopic fractional QHE physics\cite{Laughlin}
which gives rise to the chemical potential jump at $\nu =1/m$,
that the low-energy neutral edge excitations
for any $m$ are in one-to-one correspondence with those of a chiral
one-dimensional non-interacting fermion system which has
single-particle states with only one sign of velocity.
The non-interacting fermion system can in turn be mapped\cite{mahanll} to a
1D system of non-interacting chiral bosons; the
 bosons are the phonon modes of the
edge.  The same conclusion about the neutral excitations of a $\nu =1/m$ edge
can be reached by using a hydrodynamic approach\cite{wen,wenreview}
to derive a low-energy effective Hamiltonian
expressed solely in terms of 1D charge densities obtained by integrating
the 2D charge density along the coordinate perpendicular to the edge.
The Hamiltonian is quantized by invoking a commutation relation
between Fourier components of the 1D charge density:
\begin{equation}
[\rho(q),\rho(-q')]=\nu \frac{q L}{2 \pi} \delta_{q,q'}.
\label{eq:ahm2}
\end{equation}
For $\nu =1$ Eq.(~\ref{eq:ahm2}) can be derived microscopically
but for the fractional case it is assumed in order to satisfy
Eq.(~\ref{eq:ahm1}) without altering the structure of the theory.
This seemingly innocent introduction of the factor $\nu$ on
the right-hand-side of Eq.(~\ref{eq:ahm2}) is responsible for
the non-Fermi-liquid behavior predictions\cite{wen,wenreview,kanefisher}
of CLL theory.  The predictions follow from the
expression for the electron field operator,
which plays a central role in the theory, in terms of boson operators.
This expression, quoted below, is the simplest choice which
results in a field operator that satisfies the correct
commutation relation with the density operator.
For the fractional case, it has not been possible to fully
justify this expression on the basis of microscopic theory.  This situation
motivates the extensive numerical tests
reported in this article.\cite{previousnumerical}
Our findings are in complete agreement with the CLL theory.

Our numerical calculations were performed for a QHD
with a parabolic confinement potential.
These model systems\cite{macdyang,Palacios}
are of direct relevance to transport\cite{Kastner,Klein,Thomas}
and capacitance\cite{Ashoori} measurements in
quantum dots in strong magnetic fields.
Here we are interested principally in the edge excitations
of QHD's which are as large as possible in order to
model the edge excitations of macroscopic incompressible states.
In this model the single-particle state with angular momentum $m$ has
energy $ (m + 1) \Omega_0^2 \ell^2 $ where
$\Omega_0$ is the frequency characterizing the parabolic confining
potential and $\ell = (\hbar c /e B)^{1/2}$ is the magnetic length.
For a QHD the angular momentum plays the role
which would be played by the linear momentum in units of $2 \pi / L$ in a
Hall bar geometry.  For this microscopic model system the
incompressible ground states $\Psi^m_0(N)$
associated with the $\nu =1/m$ fractional QHE
occur\cite{macdedge}
for total angular momentum
$M = M_0(m,N) = m N (N-1) /2$ with $N$ being the number of particles
in the QHD. The $m=1$ incompressible state
is a single Slater determinant in which single-particle states with
$m = 0, 1, \cdots, N-1$ are occupied.

In the language appropriate to the QHD the principal predictions of the
CLL theory are the following:
i) For $M = M_0(m,N) + \Delta M$ the spectrum has a
low-energy branch with many-particle eigenenergies given by
$E = E_0 + \sum_l n_l e_l$ where
$\sum_l l n_l = \Delta M$
and $E_0$ is the energy of the
$M_0(m,N)$ incompressible state. This property is expected to be
accurately satisfied for $\Delta M < N^{1/2}$, since the
excitations are then well localized at the edge. Here $n_l$ are
the non-negative
integers which give the occupation numbers for the bosonic edge-wave angular
momentum $l$ and energy $e_l$. ii) The electron creation
operator is given by
\begin{equation}
\hat \psi^{\dagger} (\theta) = \sqrt{z} \exp
 [\hat \phi_{+}(\theta)] \exp[- \hat \phi_{-}(\theta)]
\label{eq:ahmef}
\end{equation}
where
\begin{equation}
\hat \phi_{+} (\theta) = \sum_l \sqrt{1/l \nu} [a_l^{\dagger} exp(i l \theta
)],
\label{eq:amhbf}
\end{equation}
$a_l^{\dagger}$ is a boson creation operator, $z$ is a constant which is
not fixed in the CLL theory,
$\phi_{-}(\theta) = [\phi_{+}(\theta)]^\dagger$, and for each particle number
the boson operators act on the bosonic quantum numbers of the
edge waves.

We first discuss our tests of the CLL theory predictions for the
bosonic nature of the excitation spectrum of the edge of the QHD.
Substantial arguments can be advanced in favor of this aspect of the CLL theory
predictions from microscopic theory.  In the case of the
hard-core model of electron-electron interactions, the low-energy
portion of the spectrum has no contributions from electron-electron
interactions, the bosonization follows from analytic arguments and
$e_l = l \Omega_0^2 \ell^2$.\cite{macdedge,macdjohnson}  More generally,
qualitative aspects of the bosonization of the low-energy portion
of the spectrum follow from the adiabatic
evolution of the spectrum with changing
model interactions.  For the QHD the $l =1$ single-boson state
corresponds microscopically to an excitation of the center-of-mass
of all electrons from $M_{cm}=0$ to $M_{cm}=1$.  Since the interaction
energy of the electrons is independent of the center-of-mass
state\cite{trugman,ajp} it follows $e_1= \Omega_0^2 \ell^2$,
independent of electron-electron interactions.  For $l \ne 1$
$e_l$ is interaction-dependent; to test the bosonization
for the physically realistic Coulomb interaction model we
have determined the spectrum of  finite-size QHD's by exact diagonalization
of the many-particle Hamiltonian neglecting confinement.
For parabolically confined QHD's
the eigenstates are unchanged and the subspace spectrum
at total angular momentum $M$ shifts rigidly
by $ M  \Omega_0^2 \ell^2$ when the confinement is introduced.

Figure \ref{fig1} shows the spectra of the QHD for $\Omega_0 = 0$
as a function of $M$ close to $M = M_0(3,N)$ and $M= M_0(1,N)$.
Note that the interaction energy decreases as
$M$ increases because of the decrease in average two-dimensional
electron density; in the QHD case interactions lower the boson energies because
the charge spreads out in the direction perpendicular to the edge when
the angular momentum is changed.
Fig.(~\ref{fig1}) shows that
the bosonization law for the neutral edge-wave excitation spectrum,
which is exact for the hard-core model, is still closely obeyed for
the physically realistic Coulomb interactions.
The bosonization is even more robust than would be expected {\it a priori}
since it appears to hold even where $e_l \propto l $ fails.

More extensive analyses of the spectrum are possible and have been
completed previously {\em only} for the edge excitations of a $\nu = 1 $ QHD
where exact diagonalization techniques can be applied for much larger
numbers of electrons.\cite{Stone}
A set of microscopic operators have been proposed by Stone\cite{Stone} to
generate the edge-wave spectrum in the $\nu=1$ case:
$S^\dagger_{\Delta M}= \sum_{m} ((m+\Delta M)!/m!)^{1/2}
c_{m+\Delta M}^\dagger c_{m}$ where $c_{m}^{\dagger}$ and $c_{m}$
are the electron creation and annihilation operators.  Recently,
Oaknin {\em et al.}\cite{Jacob} identified a modified set of operators,
$J^\dagger _{\Delta M}= \sum_{m} (m!/(m+\Delta M)!)^{1/2}
c_{m+\Delta M}^\dagger c_{m}$ which, when acting upon the $\nu=1$ QHD,
do not alter the center-of-mass state of the electrons.
It was demonstrated by Oaknin {\em et al.} that their operators generate the
single-boson excitations of the $\nu =1$ QHD more accurately than those
of Stone.\cite{Jacob}  More generally we find by comparing with
exact eigenstates, that for a given $\Delta M$
the excitations with large bosonic
occupation numbers are more accurately
generated by the $S^\dagger$ operators than by the $J^\dagger$ operators,
while the comparison is reversed for small bosonic occupation numbers.
In the limit $\Delta M < N^{1/2}$ the states generated by the two
sets of operators become equivalent.

In order to access large $N$ for a fractional QHD,
following Ref.\onlinecite{macdedge} we propose a set of microscopic
many-body wavefunctions for low-lying excited states of fractional QHD's:
\begin{equation}
 D^{m-1} \prod_l {(J^{\dagger}_l)^{n_l}\ } |\Psi_0^1(N)\rangle
\label{map}
\end{equation}
where $\sum_l l n_l = \Delta M$ and $D$ is the Vandermonde
determinant\cite{macdedge} Jastrow factor which relates different
Laughlin QHD states [$\Psi^m_0(N) = D^2 \Psi^{m-2}_0(N)$].
(As discussed above for some states greater accuracy at finite $N$
can be achieved on substituting $J^\dagger$ by $S^\dagger$).
These trial wave functions have the following properties: (i)
they have the appropriate value of the angular momentum;
(ii) for $\Delta M =0$ they reduce to Laughlin's approximation
for the incompressible state many-body wavefunction;
(iii) they are exact eigenstates with zero interaction energy in the case of
the hard-core model Hamiltonian; and
(iv) they are eigenstates of center-of-mass angular momentum
with eigenvalue $0$ for $\Delta M \ne 1$ and eigenvalue 1 for
$\Delta M =1$.  If the set $S^\dagger$ is used instead $J^\dagger$
the two final properties
still hold but only in the limit $\Delta M <  N^{1/2}$. Table I shows
the absolute value of the overlaps
between the trial single-boson state wavefunctions
and the exact ones for the $\nu=1/3$ QHD for up to six particles.  We
see that the excited state wavefunctions
($\Delta M \ge 2$) tend to be, if anything,
more accurate than the Laughlin trial wavefunction for the ground
state ($\Delta M = 0$). (The bracketed overlap values were calculated using
the Stone operators). It is important to realize that the direct
application of the Oaknin {\it et al.} operators to the $\nu = 1/3$
Laughlin state would not generate identical states since
$[D^{m-1},J^\dagger_{\Delta M}]\ne 0$; the resulting states would
lack properties iii) and iv) from the list above and
they are very poor approximations to the true states.
On the other hand $[D^{m-1},S^\dagger_{\Delta M}]=0$.

We turn finally to our numerical
tests of the less fully justified CLL theory expression for the
the electron field operator.  We report here only results for the squared
matrix elements connecting the ground state of the $N$ electron system with
the ground state and low-lying bosonic states of the $N+1$ particle
system at $\nu=1/3$,
$|\langle \Psi^3_{\{n_l\}} (N+1)|\psi^{\dagger}(\theta)|\Psi_0^3(N)\rangle
|^2$. The values predicted for these squared matrix elements by the CLL theory
can easily be computed from Eq.\ (\ref{eq:ahmef}).  The predicted
matrix elements for $\Delta M = 0,1,2,3,4 $ are listed in Table II
and compared with those calculated microscopically\cite{notejuanjo}
for $N=6,7$, and 8.
For each case the matrix elements have been normalized to the
ground-state-to-ground-state matrix element, $z$.\cite{note}
Since the angular momentum difference between
the $N$ and $N+1$ particle states is $M_0(3,N+1) + \Delta M - M_0(3,N)
= 3 N + \Delta M $ only the part of the electron creation operator
proportional to $c_{3N + \Delta M}^{\dagger}$ contributes to the
microscopic matrix element.  The agreement between
the CLL results for these matrix elements and the microscopic
calculations is excellent.  It appears from our extrapolation
that the CLL theory becomes exact for $N \to \infty$.
The non-Fermi-liquid power law properties predicted
by the CLL theory, depend both on the the predictions for these
matrix elements and on having
dispersionless propagation of edge modes, {\it i.e.}, $e_l = c l$.
In that case the total spectral weight at an
energy $c l$ above the chemical potential is proportional to the sum
of the squared matrix elements for $\Delta M =l$.  It follows from Eq.\
(\ref{eq:ahmef}) that this sum equals,
\begin{equation}
A_{\Delta M} = z \frac{ (\Delta M + m -1)!}{ (\Delta M)!(m-1)!}.
\label{eq:totsf}
\end{equation}
For $\Delta M \gg (m-1)$ the right-hand side of Eq.(~\ref{eq:totsf})
approaches $(\Delta M)^{m-1}/(m-1)!$.  This gives the power law dependence
of the spectral weight at low energies which enters into various physical
properties.

In conclusion, our microscopic numerical
studies strongly support the chiral Luttinger
liquid theory of the fractional Hall edge proposed by Wen and
add to the motivation for experiments which can probe the
low-energy excitations of incompressible fractional Hall states.
We thank Luis Brey, Matthew Fisher, Steve Girvin, Rudolf Haussman,
Charles Kane, Kyungsun Moon, Michael Geller,
Jacob H. Oaknin, Carlos Tejedor, and Ulrich Zuelicke for
informative and stimulating discussions.  This work has been supported by
the National Science Foundation under grant DMR-9416902.  J.J.P
acknowledges support from NATO postdoctoral research fellowship.

\begin{table}
\caption{Absolute value of the overlaps between the exact
single-boson state (see text) and the trial states
$D^2 J^\dagger_{\Delta M} |\Psi^1_0(N) \rangle $ for up to six particles
(in brackets using $S^\dagger_{\Delta M}$).
The high overlaps confirm the validity of the mapping (through $D^2$)
between the single-boson state excitations of a $\nu=1$ QHD
and those of a $\nu=1/3$ QHD.}

\begin{tabular}{|c|cccc|}

$ \Delta M $ & $N=4$  & $N=5$ & $N=6$ & \\ \hline
0&  0.9788 (0.9788) & 0.9850 (0.9850) & 0.9819 (0.9819) &\\
1&  0.9788 (0.9788) & 0.9850 (0.9850) & 0.9819 (0.9819) &\\
2&  0.9768 (0.9660) & 0.9715 (0.9647) & 0.9790 (0.9743)&\\
3&  0.9906 (0.9167) & 0.9736 (0.9296) & 0.9725 (0.9429)&\\
4&  0.9940 (0.7489) & 0.9970 (0.8373) & 0.9701 (0.8637)&\\
5&                  & 0.9862 (0.6360) & 0.9819 (0.7274)&\\
6&                  &                 & 0.9820 (0.5306)&\\
\end{tabular}
\end{table}

\begin{table}
\caption{Microscopic and CLL theory spectral weights for the
$\nu =1/3$ QHD state in units of the ground-state-to-ground-state matrix
element.
Results are shown for $N=6,7,8$ and are extrapolated to $N \to \infty$.
The bracketed CLL theory results are for $\nu=1$ where the squared matrix
elements sum to one for each $\Delta M$.  For $\nu =1/3$ each
matrix element is increased by a factor of $3^k$ where $k = \sum_l n_l$.  The
increase in the spectral weights moving away the Fermi level is due
to the increasing boson occupation numbers.}

\begin{tabular}{|c|c|cccc|c|}

$\Delta M$ & $\{n_l\}$ & \multicolumn{4}{c|}{$|\langle \Psi^3_{\{n_l\}}(N+1)|
c^\dagger_{3N+\Delta M}|\Psi^3_0(N)\rangle|^2$} & CLL theory   \\ \hline
  &  & $N=6$    & $N=7$  & $N=8$  & $N\rightarrow \infty$ & \\ \hline
0 & \{0000\}  &  1.000  &  1.000  & 1.000 & 1.000 & 1 (1)  \\ \hline
1 & \{1000\}  &  2.714  &  2.750  & 2.778 & 2.998 & 3 (1) \\ \hline
2 & \{2000\}  &  3.877  &  3.953  & 4.012 & 4.473 & 9/2 (1/2)  \\
  & \{0100\}  &  1.322  &  1.343  & 1.358 & 1.445 & 3/2 (1/2)  \\ \hline
3 & \{3000\}  &  3.877  &  3.953  & 4.012 & 4.473 & 9/2 (1/6)  \\
  & \{1100\}  &  3.913  &  3.986  & 4.041 & 4.453 & 9/2  (1/2)  \\
  & \{0010\}  &  0.939  &  0.943  & 0.946 & 0.983 & 1 (1/3) \\ \hline
4 & \{4000\}  &  3.047  &  3.088  & 3.121 & 3.360 & 27/8 (1/24)  \\
  & \{2100\}  &  6.024  &  6.131  & 6.209 & 6.710 & 27/4 (1/4)   \\
  & \{1010\}  &  2.828  &  2.852  & 2.869 & 2.966 & 3 (1/3)  \\
  & \{0200\}  &  1.048  &  1.058  & 1.064 & 1.101 & 9/8 (1/8)  \\
  & \{0001\}  &  0.830  &  0.811  & 0.797 & 0.725 & 3/4 (1/4)  \\
\end{tabular}
\end{table}

\begin{figure}
\caption{Low-lying excitation spectra of a 6-particle QHD
neglecting confinement energies.
(a) $\nu=1$ The incompressible ground state occurs at M=15.  The solid
dots show the interaction energies of the single-boson edge states, $e_l$.
($e_1=0$) The highlighted asterisk at M=19 shows the energy of the state
with $n_2 = 2$.  Its energy differs from $2 e_2$ by
3.3\%. The sequence of states which appear along horizontal lines in
this figures correspond microscopically to increasing values of the
center-of-mass angular momentum and, in the boson picture, to increasing
values of $n_1$.
(b) $\nu=1/3$ The incompressible ground state occurs at M=45.  The
similarity with the $\nu =1$ spectrum in (a) is clear despite the appearance
for $\Delta M \ge 3$ of other states representing bulk excitations.
These states are expected to move to relatively higher energies for large $N$.
If higher LL's were included in the calculation, the
same incursion of bulk excitations for values of $\Delta M$ near $N$
could occur in the $\nu=1$ case at small enough values of $\hbar \omega_c$.
In the $\nu =1/3$ case the energy of the highlighted $n_2 =2 $ state
differs from $2 e_2 $ by 0.3\%.}
\label{fig1}
\end{figure}


\begin{references}
\bibitem{notejuanjo} Here we present results obtained using
the $S^\dagger$ operators to generate the microscopic many-body states.
\bibitem{note} As has already been pointed out by Wen
this constant goes to zero in the thermodynamic limit as $N^{-1}$.
\end{references}
\end{document}